\documentstyle[epsfig,12pt,a4p]{article}

\newcommand{\pic}{\mbox{$\pi^+\pi^-\pi^+\pi^-$ }}

\newcommand{\etapipi}{\mbox{$\eta \pi^+\pi^-$ }}

\begin{document}
\begin{titlepage}
\def\footnoterule{\hrule width 1.0\columnwidth}
%\hfill  \hfill
%6\thinspace February\thinspace 1990
% \begin{center} {\large EUROPEAN ORGANIZATION FOR NUCLEAR RESEARCH}
%  \end{center}
\begin{tabbing}
put this on the right hand corner using tabbing so it looks
 and neat and in \= \kill
\> {25 November 1999}
\end{tabbing}
\bigskip
\bigskip
\begin{center}{\Large  {\bf A study of the
\etapipi channel produced in central
pp interactions at 450 GeV/c}
}\end{center}

\bigskip
\bigskip
\begin{center}{        The WA102 Collaboration
}\end{center}\bigskip
\begin{center}{
D.\thinspace Barberis$^{  4}$,
F.G.\thinspace Binon$^{   6}$,
F.E.\thinspace Close$^{  3,4}$,
K.M.\thinspace Danielsen$^{ 11}$,
S.V.\thinspace Donskov$^{  5}$,
B.C.\thinspace Earl$^{  3}$,
D.\thinspace Evans$^{  3}$,
B.R.\thinspace French$^{  4}$,
T.\thinspace Hino$^{ 12}$,
S.\thinspace Inaba$^{   8}$,
A.\thinspace Jacholkowski$^{   4}$,
T.\thinspace Jacobsen$^{  11}$,
G.V.\thinspace Khaustov$^{  5}$,
J.B.\thinspace Kinson$^{   3}$,
A.\thinspace Kirk$^{   3}$,
A.A.\thinspace Kondashov$^{  5}$,
A.A.\thinspace Lednev$^{  5}$,
V.\thinspace Lenti$^{  4}$,
I.\thinspace Minashvili$^{   7}$,
J.P.\thinspace Peigneux$^{  1}$,
V.\thinspace Romanovsky$^{   7}$,
N.\thinspace Russakovich$^{   7}$,
A.\thinspace Semenov$^{   7}$,
P.M.\thinspace Shagin$^{  5}$,
H.\thinspace Shimizu$^{ 10}$,
A.V.\thinspace Singovsky$^{ 1,5}$,
A.\thinspace Sobol$^{   5}$,
M.\thinspace Stassinaki$^{   2}$,
J.P.\thinspace Stroot$^{  6}$,
K.\thinspace Takamatsu$^{ 9}$,
T.\thinspace Tsuru$^{   8}$,
O.\thinspace Villalobos Baillie$^{   3}$,
M.F.\thinspace Votruba$^{   3}$,
Y.\thinspace Yasu$^{   8}$.
%% \end authorlist
}\end{center}

\begin{center}{\bf {{\bf Abstract}}}\end{center}

{
The reaction
$ pp \rightarrow p_{f} (\eta \pi^+\pi^-) p_{s}$
has been studied at 450 GeV/c.
There is clear evidence for an $a_2(1320) \pi$ decay mode of the
$\eta_2(1645)$ and $\eta_2(1870)$.
In addition, there is evidence for an $a_0(980)\pi$ decay mode of both
resonances and an $f_2(1270)\eta$ decay mode of the $\eta_2(1870)$.
No evidence is found for a $J^{PC}$~=~$2^{++}$ $a_2(1320)\pi$ wave.
}
\bigskip
\bigskip
\bigskip
\bigskip\begin{center}{{Submitted to Physics Letters}}
\end{center}
%\newpage
\bigskip
\bigskip
\begin{tabbing}
aba \=   \kill
% $^\dag$ \> \small
% Deceased. \\
$^1$ \> \small
LAPP-IN2P3, Annecy, France. \\
$^2$ \> \small
Athens University, Physics Department, Athens, Greece. \\
%% $^3$ \> \small
%% Bergen University, Bergen, Norway. \\
$^3$ \> \small
School of Physics and Astronomy, University of Birmingham, Birmingham, U.K. \\
$^4$ \> \small
CERN - European Organization for Nuclear Research, Geneva, Switzerland. \\
$^5$ \> \small
IHEP, Protvino, Russia. \\
$^6$ \> \small
IISN, Belgium. \\
$^7$ \> \small
JINR, Dubna, Russia. \\
$^8$ \> \small
High Energy Accelerator Research Organization (KEK), Tsukuba, Ibaraki 305-0801,
Japan. \\
$^{9}$ \> \small
Faculty of Engineering, Miyazaki University, Miyazaki 889-2192, Japan. \\
$^{10}$ \> \small
RCNP, Osaka University, Ibaraki, Osaka 567-0047, Japan. \\
$^{11}$ \> \small
Oslo University, Oslo, Norway. \\
$^{12}$ \> \small
Faculty of Science, Tohoku University, Aoba-ku, Sendai 980-8577, Japan. \\
\end{tabbing}
\end{titlepage}
\setcounter{page}{2}
\bigskip
\par
In an analysis of the
\pic final state the WA102 collaboration observed
three peaks in the mass spectrum~\cite{re:wa1024pi}.
A spin analysis showed that the peak at 1.28 GeV was due to the
$f_1(1285)$,  the peak at 1.45 GeV could be interpreted as being due to
interference between the $f_0(1370)$ and $f_0(1500)$ and
the peak at 1.9 GeV,
called the $f_{2}(1950)$, was found to have
$I^{G}J^{PC}=0^{+}2^{++}$ and decay to
$f_2(1270)\pi \pi$ and $a_2(1320)\pi$~\cite{re:wa1024pi}.
However, it was not
possible to determine whether the $f_2(1950)$ was one resonance
with two decay modes, or two resonances, or if one of the decay modes
was spurious.
\par
One of the major problems of studying the \pic final state
is the number of possible isobar decay modes that
are present. Therefore
in this paper,
in order to study the
$a_2(1320)\pi$ final state,
an analysis is presented of the
\etapipi channel.
In addition, the spin analysis of the \pic channel
showed evidence in the $J^{PC}$~=~$2^{-+}$
$a_2(1320)\pi$ wave for the $\eta_2(1645)$ and $\eta_2(1870)$.
There has been previous evidence~\cite{cbetapipi}
that the
$\eta_2(1645)$
and $\eta_2(1870)$ may decay to
$a_0(980)\pi$ and $f_2(1270)\eta$. These two decay modes will
be searched for in the present analysis.
\par
The data come from the WA102 experiment
which has been performed using the CERN Omega Spectrometer,
the layout of which is
described in ref.~\cite{WADPT}.
The selection of
the reaction
\begin{equation}
pp \rightarrow p_{f} (\eta\pi^+\pi^-) p_{s}
\label{eq:c}
\end{equation}
where the subscripts $f$ and $s$ indicate the
fastest and slowest particles in the laboratory respectively,
has been described in ref.~\cite{f1pap}.
The $\eta$ has been observed decaying to $\gamma \gamma$ and
$\pi^+\pi^-\pi^0$.
Fig.~\ref{fi:1}a) and \ref{fi:1}b) show the
\etapipi mass spectra for the decays
$\eta \rightarrow \gamma \gamma$ and
$\eta \rightarrow \pi^+\pi^-\pi^0$ respectively.
The mass spectra are dominated by the $\eta^\prime$ and $f_1(1285)$.
\par
In this current paper a
spin-parity analysis of the
\etapipi
channel
is presented for the mass interval 1.0 to 2.0 GeV
using an isobar model~\cite{re:wa914pi}.
Assuming that
only angular momenta up to 2 contribute,
the intermediate
states considered are $a_0(980) \pi$, $\sigma \eta$, $f_0(980) \pi$,
$a_2(1320)\pi$, and $f_2(1270) \eta$.
$\sigma$ stands for the low mass $\pi\pi$ S-wave amplitude
squared~\cite{re:zbugg}.
The amplitudes have been calculated in the spin-orbit (LS)
scheme using spherical
harmonics.
In order to perform a spin parity analysis the
Log Likelihood function, ${\cal L}_{j}=\sum_{i}\log P_{j}(i)$,
is defined by combining the probabilities of all events in 40 MeV
\etapipi mass bins from 1.0 to 2.0 GeV.
The incoherent sum of various
event fractions $a_{j}$ is calculated
so as to include more than one wave in the fit,
using the form:
\begin{equation}
{\cal L}=\sum_{i}\log \left(\sum_{j}a_{j}P_{j}(i) +
(1-\sum_{j}a_{j})\right)
\end{equation}
where the term
$(1-\sum_{j}a_{j})$ represents the phase space background.
This background term is used to account for
the background below the $\eta$
(which is 10 \% for the $\gamma \gamma$ decay mode
and 15~\% for the $\pi^+\pi^-\pi^0$ decay mode), \etapipi three body decays
and decay modes not parameterised in the fit.
The negative Log Likelihood function ($-{\cal L} $) is then minimised using
MINUIT~\cite{re:MINUIT}.
Different combinations of waves and isobars have been tried and
insignificant contributions have been removed from the final fit.
\par
The spin analysis has been performed independently
for the two $\eta$ decay modes.
As was shown in the previous analysis~\cite{f1pap},
for both decay modes
and
for M(\etapipi)~$\le$ 1.5~GeV
the only wave required in the fit is the $J^{PC}$~=~$1^{++}$ $a_0(980)\pi$
wave with spin projection $|J_Z|$~=~1.
No $J^{PC}$~=~$0^{-+}$ $a_0(980)\pi$ or any $\sigma \eta$ waves
are required in the fit.
Fig.~\ref{fi:2}a) and ~\ref{fi:3}a) show the
$J^{PC}$~=~$1^{++}$ $a_0(980)\pi$ wave
where the $f_1(1285)$ and a shoulder
at 1.4~GeV can be seen.
Superimposed on the waves is the result of the fit used in
ref.~\cite{f1pap} which
uses a K matrix formalism
including poles to describe the interference between the
$f_1(1285)$ and the $f_1(1420)$.
As can be seen from
fig.~\ref{fi:2}a) and \ref{fi:3}a)
the parameterisation describes well the
$J^{PC}$~=~$1^{++}$ $a_0(980)\pi$ wave.
\par
For M(\etapipi)~$\ge$ 1.5~GeV
only waves with $J^{PC}$~=~$2^{-+}$ and $|J_Z|$~=~1 are
required in the fit.
In contrast to what was found in the analysis of the \pic final
state~\cite{re:wa1024pi}
there is no evidence for any
$J^{PC}$~=~$2^{++}$ $a_2(1320)\pi$ wave.
The largest change in Log Likelihood comes from the
addition of the
$J^{PC}$~=~$2^{-+}$ $a_2(1320)\pi$ wave with $|J_Z|$~=~1 which
yields a Likelihood difference $\Delta{\cal L}$~=~562 and 203
for the
$\eta \rightarrow \gamma \gamma$ and
$\eta \rightarrow \pi^+\pi^-\pi^0$ decays respectively and are shown in
fig.~\ref{fi:2}b) and \ref{fi:3}b).
\par
As was observed in the case of the \pic channel~\cite{re:wa1024pi}, the
$J^{PC}$~=~$2^{-+}$ $a_2(1320)\pi$ wave
is consistent with being due to two resonances,
the $\eta_2(1645)$ and the $\eta_2(1870)$.
Superimposed on
figs.~\ref{fi:2}b) and \ref{fi:3}b) is the result of a fit
using a single channel K matrix formalism~\cite{KMATRIX}
with two resonances to describe the $\eta_2(1645)$
and $\eta_2(1870)$.
The masses and widths determined for each resonance
and each $\eta$ decay mode are given in table~\ref{ta:a}.
The parameters found are consistent for the two decay modes
and with the PDG~\cite{PDG98} values for these resonances.
An alternative fit has been performed using two interfering
Breit-Wigners. The parameters presented
include not only the statistical error but also
the systematic error, added in quadrature,
representing the difference in the two fits.
\par
The addition of the
$J^{PC}$~=~$2^{-+}$ $a_0(980)\pi$ wave with $|J_Z|$~=~1
yields a Likelihood difference $\Delta{\cal L}$~=~66 and 23
for the two $\eta$ decay modes respectively and the waves are shown in
figs.~\ref{fi:2}c) and \ref{fi:3}c).
Superimposed
is the result of a fit
using the parameters for the $\eta_2(1645)$
and the $\eta_2(1870)$ determined from the fit
to the $a_2(1320)\pi$ final state.
A good description of the data is found.
The branching ratio of the $\eta_2(1645)$ and
$\eta_2(1870)$ to $a_2(1320)\pi/a_0(980)\pi$
in the $\eta\pi \pi$ final state
can be determined.
Neglecting unseen decay modes
the branching ratio
of $\eta_2(1645)$
to $a_2(1320)\pi$/$a_0(980)\pi$~=~2.3~$\pm$~0.4 and
{}~2.1~$\pm$~0.5 for
the decays
$\eta \rightarrow \gamma \gamma$ and
$\eta \rightarrow \pi^+\pi^-\pi^0$ respectively.
The
branching ratio
of $\eta_2(1870)$
to $a_2(1320)\pi$/$a_0(980)\pi$~=~5.0~$\pm$~1.6 and
{}~6.0~$\pm$~1.9 respectively.
\par
Correcting for the unseen $a_2(1320)$ decay modes using the PDG~\cite{PDG98}
branching ratio and using the branching ratio for the $a_0(980)$ to
$\eta\pi$ determined by this experiment~\cite{f1pap}
of 0.86~$\pm$~0.10 and taking the average of the
two $\eta$ decay modes the branching ratio
to $a_2(1320)\pi$/$a_0(980)\pi$
for the $\eta_2(1645)$ is 13.0~$\pm$~2.7 and
for the $\eta_2(1870)$ is 32.6~$\pm$~12.6.
\par
The addition of the
$J^{PC}$~=~$2^{-+}$ $f_2(1270)\eta$ wave with $|J_Z|$~=~1
yields a
Likelihood difference $\Delta{\cal L}$~=~42 and 12 and the waves
are shown in
figs.~\ref{fi:2}d) and \ref{fi:3}d).
The $f_2(1270)\eta$ wave shows little evidence
for the $\eta_2(1645)$.
Superimposed
is the result of a fit
using the parameters for the
$\eta_2(1870)$ determined from the fit
to the $a_2(1320)\pi$ final state.
A satisfactory description of the data is found.
Correcting for the unseen $a_2(1320)$ and $f_2(1270)$ decay modes
the branching ratio of the
$\eta_2(1870)$ to $a_2(1320)\pi/f_2(1270)\eta$
has been determined to be
% 38.5~$\pm$~14.4 and
% ~56.0~$\pm$~33.4 for
19.2~$\pm$~7.2 and
{}~27.6~$\pm$~16.4 for
the two $\eta$ decay modes.
The addition of the
$J^{PC}$~=~$2^{-+}$ $f_0(980)\eta$ or
$J^{PC}$~=~$2^{-+}$ $\sigma\eta$ waves produces no
significant change in the Likelihood and hence they have
been excluded from the fit.
The resulting background term is found to be smooth and structureless
and corresponds to $\approx$~40~\% of the channel.
\par
In previous analyses it has been observed that
when the centrally produced system has been analysed
as a function of the parameter $dP_T$, which is the difference
in the transverse momentum vectors of the two exchange
particles~\cite{WADPT,closeak},
all the undisputed
$ q \overline q $ states
(i.e. $\eta$, $\eta^{\prime}$, $f_1(1285)$ etc.)
are suppressed as $dP_T$ goes to zero,
whereas the glueball candidates
$f_0(1500)$, $f_0(1710)$ and $f_2(1950)$ are prominent~\cite{memoriam}.
In order to calculate the contribution of each resonance as a function
of $dP_T$, the waves have been fitted
in three $dP_T$ intervals with
the parameters of the resonances fixed to those obtained from the
fits to the total data.
Table~\ref{ta:dpt} gives the percentage of each resonance
in three $dP_T$ intervals together with the ratio of the number of events
for $dP_T$ $<$ 0.2 GeV to
the number of events
for $dP_T$ $>$ 0.5 GeV for each resonance considered.
These distributions are similar to what have been
observed for other $q \overline q$ states~\cite{memoriam}.
\par
In addition, an interesting effect has been observed in
the azimuthal angle $\phi$ which is defined as the angle between the $p_T$
vectors of the two outgoing protons.
For the resonances
studied to date which are compatible with
being produced by DPE,
the data~\cite{phiangpap}.
are consistent with the Pomeron
transforming like a non-conserved vector current~\cite{clschul}.
In order to determine the
$\phi$ dependence for the resonances observed,
a spin analysis has been performed in the \etapipi channel
in four different $\phi$ intervals each
of 45 degrees.
The results are shown in fig.~\ref{fi:4}a) and b) for the
$\eta_2(1645)$ and $\eta_2(1870)$ respectively.
\par
In order to determine the
four momentum transfer dependence ($|t|$) of the
resonances observed
in the \etapipi channel
the waves have been fitted in 0.1 GeV$^2$ bins
of $|t|$
with the parameters of the resonances fixed to those obtained from the
fits to the total data.
Fig.~\ref{fi:4}c) and d) show the four momentum transfer from
one of the proton vertices for the
$\eta_2(1645)$ and $\eta_2(1870)$ respectively.
The distributions
cannot be fitted with a single
exponential.
Instead they have been fitted to the form
\[
\frac{d\sigma}{dt} = \alpha e^{-b_1t} + \beta t e^{-b_2t}
\]
The parameters resulting from the fit are
given in table~\ref{ta:c}.
\par
After correcting for
geometrical acceptances, detector efficiencies,
losses due to cuts,
and unseen decay modes of the $a_2(1320)$,
the cross-section for
the $\eta_2(1645)$ and $\eta_2(1870)$ decaying to $a_2(1320)\pi$
at $\sqrt s$~=~29.1~GeV in the
$x_F$ interval
$|x_F| \leq 0.2$ has been determined to be
$\sigma(\eta_2(1645))$~=~1664~$\pm$~149~nb,
and $\sigma(\eta_2(1870))$~=~1845~$\pm$~183~nb.
\par
A Monte Carlo simulation has been performed for the production
of a $J^{PC}$~=~$2^{-+}$ state with spin projection $|J_Z|$~=1
via the exchange of two non-conserved vector currents
using the model of Close and Schuler~\cite{clschul}
discussed in ref.~\cite{phiangpap}. The prediction
of this model is found to be in qualitative agreement
with the observed $\phi$, $dP_T$ and $t$ distributions
of the $\eta_2(1645)$ and $\eta_2(1870)$.
This will be presented in a later publication.
\par
In summary,
there is evidence for an $a_2(1320) \pi$ decay mode of the
$\eta_2(1645)$ and $\eta_2(1870)$ in the \etapipi
final state.
In addition, there is evidence for an $a_0(980)\pi$ decay mode of both
resonances and possibly a $f_2(1270)\eta$ decay mode of the $\eta_2(1870)$.
There is no evidence for any $J^{PC}$~=~$2^{++}$ $a_2(1320)\pi$ wave,
in particular no evidence for the decay
$f_2(1950)$~$\rightarrow$~$a_2(1320)\pi$.
\begin{center}
{\bf Acknowledgements}
\end{center}
\par
This work is supported, in part, by grants from
the British Particle Physics and Astronomy Research Council,
the British Royal Society,
the Ministry of Education, Science, Sports and Culture of Japan
(grants no. 07044098 and 1004100), the French Programme International
de Cooperation Scientifique (grant no. 576)
and
the Russian Foundation for Basic Research
(grants 96-15-96633 and 98-02-22032).

\newpage
{ \large \bf Tables \rm}
\begin{table}[h]
\caption{Parameters of the $\eta_2(1645)$ and $\eta_2(1870)$}
\label{ta:a}
\vspace{1in}
\begin{center}
\begin{tabular}{|c|c|c|c|} \hline
  & & & \\
 Resonance &Final state&Mass (MeV) & Width (MeV) \\
 & & &  \\
 & & &  \\ \hline
 & & &  \\
$\eta_{2}(1645)$ & $\eta \pi\pi $ &1605 $\pm$ 12 &188 $\pm$ 22  \\
 &$\eta \rightarrow \gamma \gamma $ & &  \\
$\eta_{2}(1645)$ & $\eta \pi\pi $ &1619 $\pm$ 11 &179 $\pm$ 28  \\
 &$\eta \rightarrow \pi^+\pi^-\pi^0 $ & &  \\
 & & &  \\ \hline
 & & &  \\
$\eta_{2}(1870)$ & $\eta \pi\pi $ &1841 $\pm$ 18 &249 $\pm$ 30  \\
 &$\eta \rightarrow \gamma \gamma $ & &  \\
$\eta_{2}(1870)$ & $\eta \pi\pi $ &1831 $\pm$ 16 &219 $\pm$ 33  \\
 &$\eta \rightarrow \pi^+\pi^-\pi^0 $ & &  \\
 & & &  \\ \hline
\end{tabular}
\end{center}
\end{table}
\newpage
\begin{table}[h]
\caption{Production of the resonances as a function of $dP_T$
expressed as a percentage of their total contribution and the
ratio (R) of events produced at $dP_T$~$\leq$~0.2~GeV to the events
produced at $dP_T$~$\geq$~0.5~GeV.}
\label{ta:dpt}
\vspace{1in}
\begin{center}
\begin{tabular}{|c|c|c|c|c|} \hline
 & & & & \\
 &$dP_T$$\leq$0.2 GeV & 0.2$\leq$$dP_T$$\leq$0.5 GeV &$dP_T$$\geq$0.5 GeV &
$R=\frac{dP_T \leq 0.2 GeV}{dP_T\geq 0.5 GeV}$\\
 & & & & \\ \hline
 & & & & \\
$\eta_2(1645)$  &8.9 $\pm$ 1.1  & 32.2 $\pm$ 3.0 &58.9 $\pm$ 5.2 &
0.15~$\pm$~0.02\\
 & & & & \\ \hline
 & & & & \\
$\eta_2(1870)$  &8.2 $\pm$ 1.0  & 28.6 $\pm$ 2.8 &63.2 $\pm$ 5.6 &
0.13~$\pm$~0.02\\
 & & & & \\ \hline
\end{tabular}
\end{center}
\end{table}
\newpage
\begin{table}[h]
\caption{The slope parameters from a fit to the
$|t|$ distributions of the form
$\frac{d\sigma}{dt} = \alpha e^{-b_1t} + \beta t e^{-b_2t}$.}
\label{ta:c}
\vspace{0.3in}
\begin{center}
\begin{tabular}{|c|c|c|c|c|} \hline
  &&&&  \\
 & $\alpha$ & $b_1$ & $\beta$ & $b_2$\\
&&GeV$^{-2}$&&GeV$^{-2}$ \\
   &&&& \\ \hline
  &&&& \\
$\eta_2(1645)$ & $0.4\pm 0.1$& $6.4 \pm 2.0$& $2.6 \pm 0.9$& $7.3 \pm 1.3$ \\
  &&&& \\
$\eta_2(1870)$ & $0.3\pm 0.1$& $5.9 \pm 3.5$& $4.3 \pm 1.5$& $8.3 \pm 2.0$ \\
   &&&& \\ \hline
\end{tabular}
\end{center}
\end{table}
\clearpage
{ \large \bf Figures \rm}
\begin{figure}[h]
\caption{
The \etapipi mass spectrum for a) $\eta \rightarrow \gamma \gamma$ and
b) $\eta \rightarrow \pi^+\pi^-\pi^0$.
}
\label{fi:1}
\end{figure}
\begin{figure}[h]
\caption{The \etapipi mass spectrum for the decay
$\eta \rightarrow \gamma \gamma$.
a) The $J^{PC}~=~1^{++}$~$a_0(980)\pi$ wave,
b) the $J^{PC}~=~2^{-+}$~$a_2(1320)\pi$ wave,
c) the $J^{PC}~=~2^{-+}$~$a_0(980)\pi$ wave and
d) the $J^{PC}~=~2^{-+}$~$f_2(1270)\pi$ wave.
}
\label{fi:2}
\end{figure}
\begin{figure}[h]
\caption{The \etapipi mass spectrum for the decay
$\eta \rightarrow \pi^+ \pi^- \pi^0$.
a) The $J^{PC}~=~1^{++}$~$a_0(980)\pi$ wave,
b) the $J^{PC}~=~2^{-+}$~$a_2(1320)\pi$ wave,
c) the $J^{PC}~=~2^{-+}$~$a_0(980)\pi$ wave and
d) the $J^{PC}~=~2^{-+}$~$f_2(1270)\pi$ wave.
}
\label{fi:3}
\end{figure}
\begin{figure}[h]
\caption{
The azimuthal angle ($\phi$) between the two outgoing protons
for a) the $\eta_2(1645)$ and d) the $\eta_2(1870)$.
The four momentum transfer squared ($|t|$) from one of the proton
vertices for c) the $\eta_2(1645)$ and d) the $\eta_2(1870)$ with
fits described in the text.
}
\label{fi:4}
\end{figure}
\begin{center}
\epsfig{figure=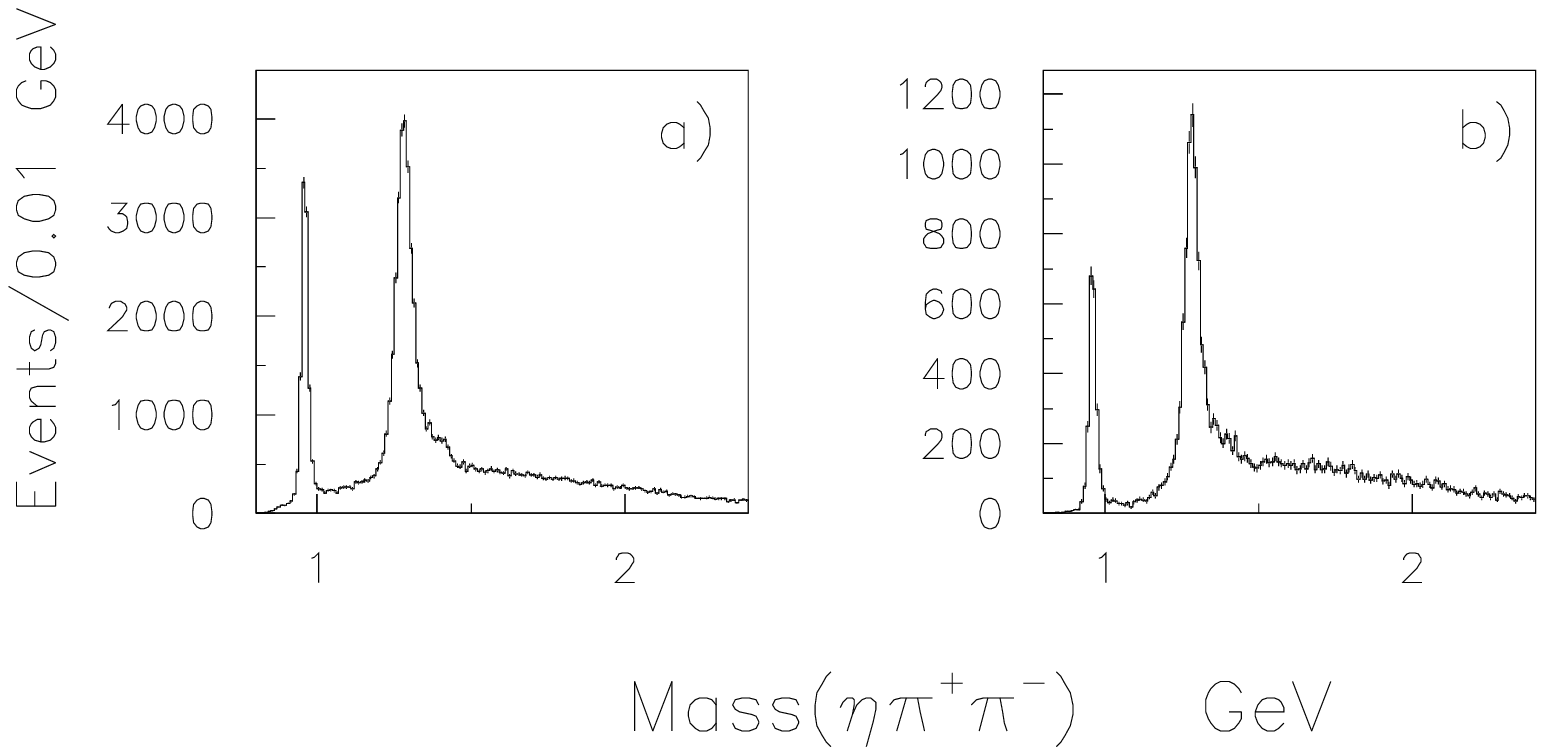,height=22cm,width=17cm}
\end{center}
\begin{center} {Figure 1} \end{center}
\newpage
\begin{center}
\epsfig{figure=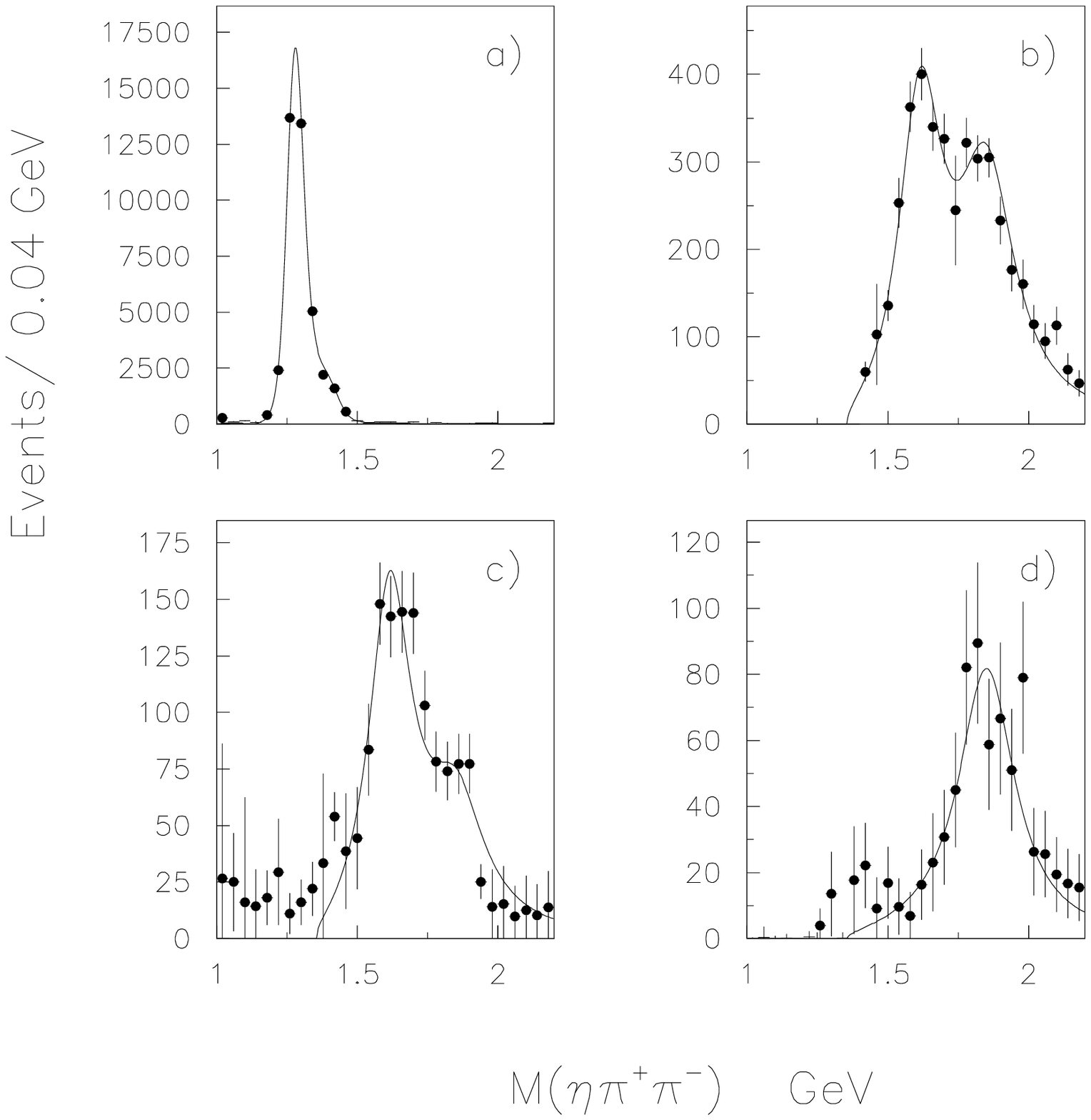,height=22cm,width=17cm}
\end{center}
\begin{center} {Figure 2} \end{center}
\newpage
\begin{center}
\epsfig{figure=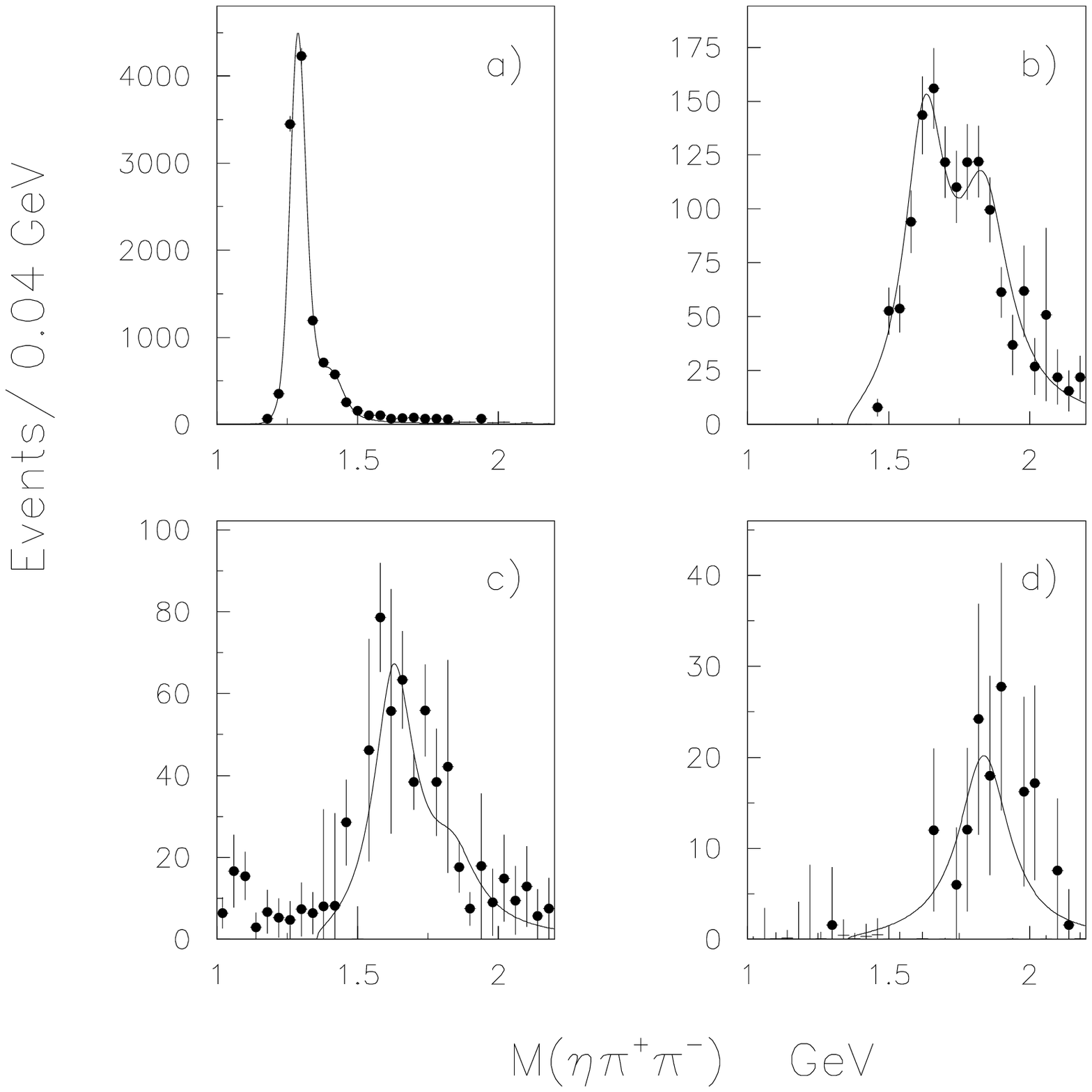,height=22cm,width=17cm}
\end{center}
\begin{center} {Figure 3} \end{center}
\newpage
\begin{center}
\epsfig{figure=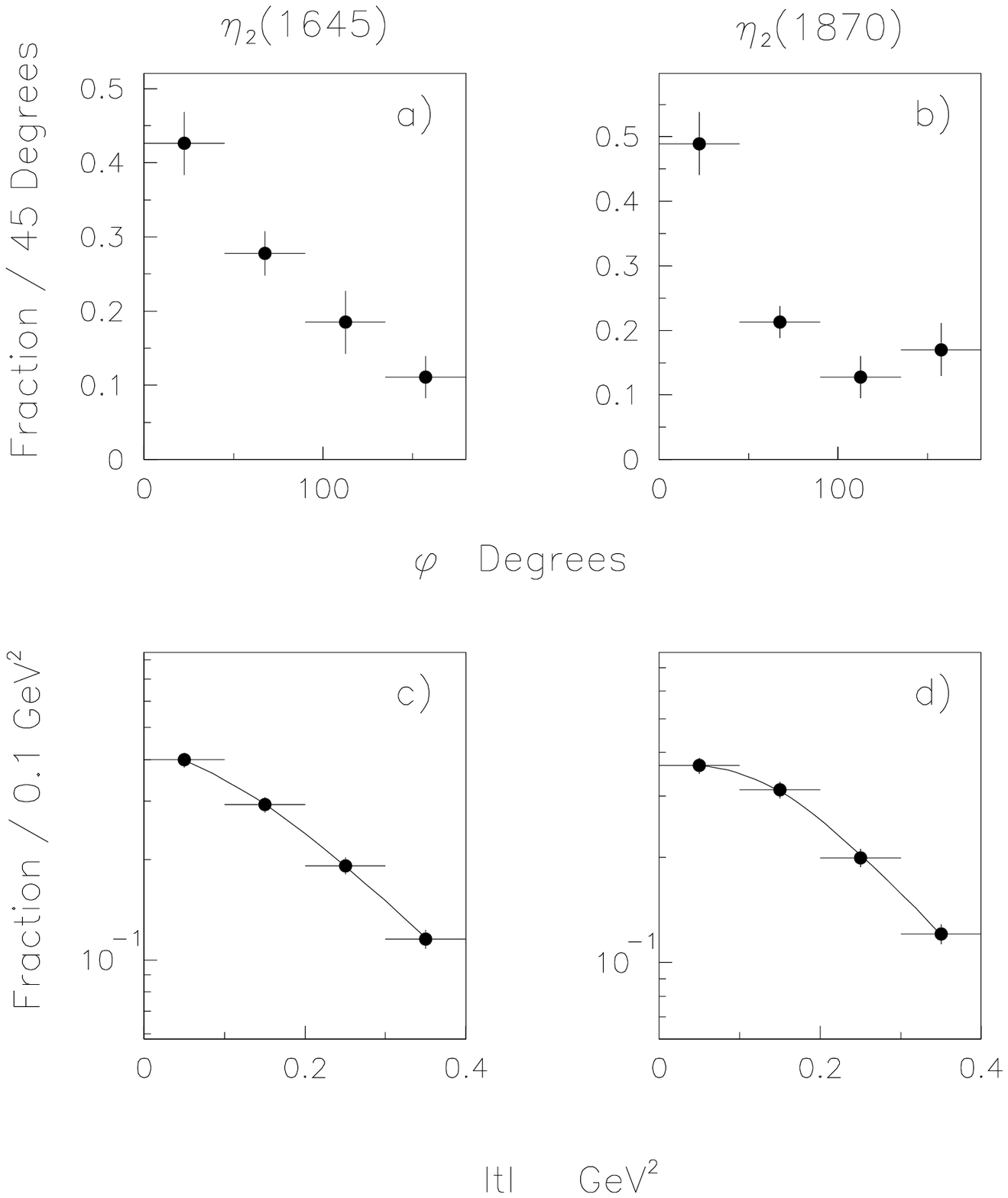,height=22cm,width=17cm}
\end{center}
\begin{center} {Figure 4} \end{center}

\bigskip
\newpage
\begin{thebibliography}{99}
\bibitem{re:wa1024pi}
D. Barberis {\em et al.,} Phys. Lett. {\bf B413 } \rm (1997) 217.
\bibitem{cbetapipi}
K. Karch  {\em et al.,} Zeit. Phys {\bf C54} (1992) 33;\\
B. Adomeit {\em et al.,} Zeit. Phys {\bf C71} (1996) 227.
\bibitem{WADPT}
D. Barberis {\em et al.,} Phys. Lett. {\bf B397 } \rm (1997) 339.
\bibitem{f1pap}
D. Barberis {\em et al.,} Phys. Lett. {\bf B440 } \rm (1998) 225.
\bibitem{re:wa914pi}
S. Abatzis {\em et al.,} Phys. Lett. {\bf B324 } \rm (1994) 509.
\bibitem{re:zbugg}
B. S. Zou and D. V. Bugg, Phys. Rev. {\bf D48} (1993) R3948; \\
K. L. Au, D. Morgan and M. R. Pennington, Phys. Rev. {\bf D35 } \rm (1987)
1633; \\
D. Barberis {\em et al.,} In preparation.
\bibitem{re:MINUIT}
F. James and M. Roos, MINUIT Computer Physics Communications
{\bf 10 } \rm (1975) 343; CERN-D506 (1989).
\bibitem{PDG98}
Particle Data Group, European Physical Journal {\bf C3} (1998) 1.
\bibitem{KMATRIX}
S.U. Chung {\em et al.,} Ann. d. Physik. {\bf 4} (1995) 404.
\bibitem{closeak}
F.E. Close and A. Kirk, Phys. Lett. {\bf B397 } \rm (1997) 333.
\bibitem{memoriam}
A. Kirk, Yad. Fiz. {\bf 62} (1999) 439.
\bibitem{phiangpap}
D. Barberis {\em et al.,} hep-ex/9909013 To be published in Phys. Lett.
\bibitem{clschul}
F.E. Close and G. Schuler, Phys. Lett. {\bf B464 } \rm (1999) 279.
\end{thebibliography}
\end{document}